\documentclass[11pt]{article}
\usepackage{epsfig}
\usepackage{amssymb}
\usepackage{amsthm}
\usepackage{amscd}
\usepackage[T1]{fontenc}
\usepackage{ae,aecompl}
\usepackage{pslatex}
\usepackage{amsfonts}
\usepackage{amsmath}
\usepackage{graphicx}
\usepackage{color}
\usepackage{latexsym}  
\usepackage{url}
\usepackage{setspace}

\newtheorem{thm}{Theorem}

\newtheorem{conj}[thm]{Conjecture}

\newtheorem{ex}[thm]{Example}

\newtheorem{cor}[thm]{Corollary}
\newtheorem{prop}[thm]{Proposition}

\begin{document}

\title{Tropical Geometry of Statistical Models}

\author{Lior Pachter and Bernd Sturmfels \\
Department of Mathematics, University of California, Berkeley, CA
94720}

\maketitle

\begin{abstract}
This paper presents a unified mathematical framework
for inference in graphical models, building on the
observation that graphical models are algebraic varieties.
 From this geometric viewpoint,
observations generated from a model are coordinates
of a point in the variety, and the sum-product algorithm is
an efficient tool for evaluating specific coordinates.
The question addressed here is how the solutions
to various inference problems depend on the model parameters.
The proposed answer is expressed in terms of
  tropical algebraic geometry. A key role is played by the
  Newton polytope of a statistical model. Our results are applied
  to  the hidden Markov model and to the
general Markov model on a binary tree.
\end{abstract}


\section{Algebraic Statistics, Tropical Geometry, and Inference}

This paper presents a unified mathematical framework for probabilistic 
inference
with statistical models, such as graphical models. Our approach is 
summarized as follows:

\textbf {(a) Statistical models are algebraic varieties.}

\textbf {(b) Every algebraic variety can be tropicalized.}

\textbf {(c) Tropicalized statistical models are fundamental for 
parametric inference.}

By a \emph{statistical model} we mean
a family of joint probability distributions for
a collection of discrete random variables
${\bf Y} = \{Y_1,\ldots,Y_n\}$. Thesis (a) states that many
families of interest can be characterized
by polynomials in the joint probabilities
$ \, p_{\sigma_1  \cdots \sigma_n} \, = \, {\rm Prob}(Y_1 = \sigma_1, 
\ldots, Y_n = \sigma_n )$.
The emerging field of algebraic statistics
\cite{Garcia:03, Pistone:01}
offers algorithms for this polynomial representation.

\emph{Tropicalization} means
replacing the arithmetic operations $(+,\times)$
by the operations $({\rm min},+)$. This process
captures the essence of what happens when the
joint probabilities $ \, p_{\sigma_1  \cdots \sigma_n} \, $ are replaced
by their logarithms.
The tropicalization of an algebraic variety is a
piecewise-linear set which enjoys many
features familiar from algebraic geometry \cite{Develin:03, 
Richter-Gebert:03}.
In particular, the tropicalization of
  a  statistical model is a piecewise-linear set
in the space with logarithmic coordinates
$\, - {\rm log}(p_{\sigma_1 \cdots \sigma_n}) $.

Thesis (c) states that
tropical algebraic geometry of statistical models
is fundamental in analyzing the behavior of inference algorithms
under the variation of model parameters.
By \emph{inference} we mean the evaluation of one or more coordinates 
of a single point on the algebraic variety, in either  $(+,\times)$ or $({\rm min},+)$ arithmetic.
This is the standard notion of inference used for
graphical models in statistical learning theory \cite{Jordan:02},
but it differs from
other (more classical) notions of inference in  mathematical statistics.
By {\em parametric inference} we mean the
analysis of the dependence of inference on parameters.

To give a more concrete discussion of parametric inference
it is useful to focus on directed graphical models. A \emph{directed graphical model}
(or \emph{Bayesian network})
is a finite directed acyclic graph $G$ with two kinds of vertices, {\em 
observed variables} ${\bf Y} = \{Y_1,\ldots,Y_n\}$ and
{\em hidden variables} ${\bf X} = \{ X_1,\ldots,X_m\}$, where each edge 
is
labeled by a transition matrix whose entries are linear forms
in some parameters.
  The rules of discrete probability express the observed
  probabilities $p_{\sigma_1 \cdots \sigma_n}$
as polynomials of degree $E$ in the parameters,
where $E$ is the number of edges of $G$.
The polynomials parametrize
the graphical model as an algebraic variety.

The two standard types of inference questions for graphical models are:
\begin{enumerate}
\item[1.] the calculation of {\em marginal probabilities}:
\[ p_{\sigma_1 \cdots \sigma_n}
\quad = \quad
\sum_{h_1,\ldots,h_m} 
{\rm Prob}(X_1  = h_1, \ldots,X_m = h_m,
Y_1 = \sigma_1,\ldots,Y_n = \sigma_n) ,\]
\item[2.] the calculation of {\em maximum a posteriori (MAP)} log
probabilities:
\[ \delta_{\sigma_1 \cdots \sigma_n}
\quad = \quad
  \min_{h_1,\ldots,h_m} 
 -{\rm log} \left( {\rm Prob}(X_1  = h_1, \ldots,X_m = h_m,
Y_1 = \sigma_1,\ldots,Y_n = \sigma_n) \right) , \]
\end{enumerate}
where the $h_i$ range over all the possible assignments for the
hidden random  variables $X_i$.
Together, these two primitives can be used to effectively solve a range 
of other inference problems, including the calculation
of conditional probabilities and other quantities of interest.
The key to inference in graphical models is the
{\em sum-product algorithm} \cite{Kschischang:01}
(also known as the \emph{generalized distributive law} \cite{Aji:00}).
This polynomial-time algorithm is used,
both in ordinary arithmetic $(+,\times)$
and in tropical arithmetic $({\rm min},+)$,
to {\em efficiently} solve Problems 1 and 2.
For more background on the sum-product algorithm,
and for connections to message passing and
the junction tree algorithm see \cite{Jordan:02}.

Although the sum-product algorithm
provides efficient solutions to the basic inference problems 1 and 2,
it only applies to one coordinate $p_{\sigma_1 \cdots \sigma_n}$
of one distribution at a time. What we are interested in
are the {\em parametric} versions of the inference problems.
They can be phrased as follows:
\begin{enumerate}
\item[3.] Find all parameters for a model which result in the same 
values for all $p_{\sigma_1,\cdots,\sigma_n}$.
\item[4.] Given observations ${\bf Y}={\bf \sigma}$ and
hidden data  ${\bf X} = {\bf h}$, identify all
parameters such that ${\bf h}$ is the most likely
explanation for the observations ${\bf \sigma}$.
\end{enumerate}
As we will see, the following {\em modeling}
questions are fundamentally related to Problems 3 and 4:
\begin{enumerate}
\item[5.] Which (parameter independent) relations on the
probabilities $p_{\sigma_1 \cdots \sigma_n}$ does the model imply?
\item[6.] Describe the tropicalization of the variety corresponding to
a graphical model.
\end{enumerate}

Problem 5 asks for the ideal of \emph{polynomial invariants} of a 
statistical model \cite{Garcia:03}.
Invariants have been investigated in
phylogenetics \cite{Allman:03, Cavender:87} where they
can help to identify good trees for aligned DNA sequences.

The primary goal of our study is to give a
practical answer to question 4 for graphical models.
Our main algorithmic result is an efficient procedure
for parametric inference
that can be viewed as a polytopal
analog of the sum-product algorithm.
The efficiency is based on the complexity  estimates
for Newton polytopes which we derive in Section 4.
The resulting {\em polytope propagation algorithm}
is applied to problems in biological sequence analysis
in the companion paper \cite{Pachter:04}.

The mathematics to be developed
in Sections 3 and 4 is of independent
interest. It also furnishes new tools for
parametric inference (Problems 
3 and 4)
and parametric modeling  (Problems 5 and 6) which are
applicable to  a wide range of statistical problems.
We demonstrate this by analyzing  the hidden Markov model (HMM)
and the general Markov model on a binary tree, in Sections 2
and 5 respectively.

\section{Algebraic Representation of Hidden Markov Models}

A graphical model is an algebraic variety which is presented as the
image of a highly structured polynomial map
$\, f \, : \, {\bf R}^d \rightarrow {\bf R}^m$.
Here $\, {\bf R}^d \,$ is the space whose coordinates are the model
parameters $s_1,\ldots,s_d$ and ${\bf R}^m$ is
the space whose coordinates $p_\sigma = p_{\sigma_1 \cdots
\sigma_n}$ are the joint probabilities for the observed
random variables.
In applications, the integer $m$ is much larger than the integer $d$,
in fact; it is so large that one can only look at one
coordinate $p_\sigma$ at a time. Each coordinate $f_\sigma =
f_\sigma(s_1,\ldots,s_d)$ of the map $f$ is
a polynomial function in $s_1,\ldots,s_d$. The efficient  evaluation
of these functions relies on
the sum-product algorithm. Here we study the
(parametric) inference and modeling problems in the familiar
context of the {\em hidden Markov model} (HMM).

A discrete HMM has $n$
observed states $Y_1,\ldots,Y_n$ taking on $l$ possible values, and $n$ 
hidden states $X_1,\ldots,X_n$ taking on $k$ possible values.
The HMM can be characterized by the following conditional
independence statements for  $i = 1 , \ldots,n$:
\begin{eqnarray*} & p(X_i \, | \,X_1,X_2,\ldots,X_{i-1}) \quad
= \quad  p(X_i \,| \, X_{i-1}),
\\& p(Y_i \, |\, X_1,\ldots,X_i,Y_1,\ldots,Y_{i-1})\quad =
\quad p(Y_i \,|\, X_i). \end{eqnarray*}
We consider the  homogeneous model with uniform
  initial distribution, where all transitions $\, X_i \rightarrow X_{i+1}\,$ are given
  by the same $k \times k$-matrix $S = (s_{ij})$ and all transitions $\,X_i \rightarrow Y_i \,$ are given by the same
$k \times l$-matrix $T = (t_{ij})$. Throughout our discussion we
disregard for simplicity the usual probabilistic hypothesis
that $S$ and $T$ are non-negative and all row sums are $1$.

\begin{prop} \label{hmmprop}
The hidden Markov model is the image of
a map $f :  {\bf R}^d \rightarrow {\bf R}^{l^n}$, where
$d = k(k+l)$ and each coordinate of $f$ is a
bi-homogeneous polynomial of degree $n-1$ in $S$
and degree $n$ in $T$.
\end{prop}

Problem 3 is to compute the fibers of the map $f$.
In statistics, this is called \emph{parameter identification}.
We use the term {\em coordinate polynomials} for 
the polynomials $f_\sigma$ that are coordinates of the map $f$.

\begin{figure}[ht]
\begin{center}
  \includegraphics[scale=0.7]{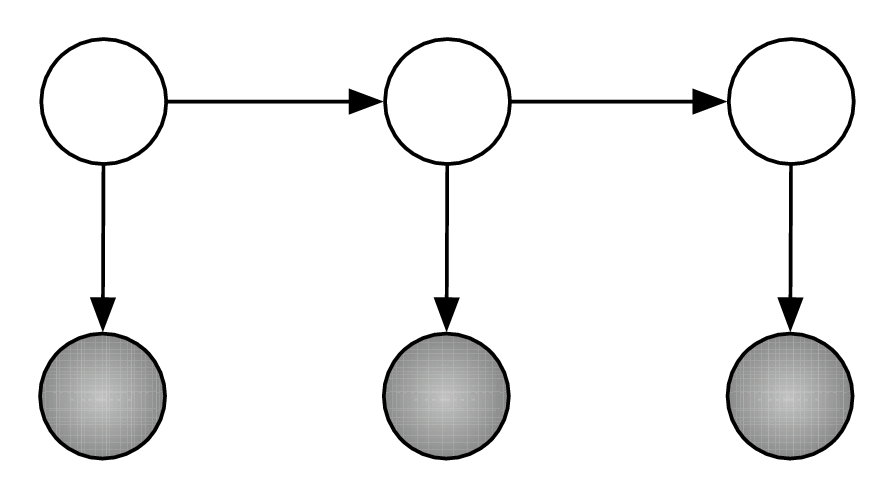}
  \end{center}
\caption{The hidden Markov model of length three.}
     \label{fig:HMM}\end{figure}

Our running example in this section is the case $n = 3$
with binary random variables $(k=l=2)$. The graph of this model
is drawn in Figure \ref{fig:HMM}.
The shaded nodes are the observed random variables.

Here the parameter space is ${\bf R}^8$ with coordinates
$ s_{00},s_{01},s_{10},s_{11},t_{00},t_{01},t_{10},t_{11} $,
and it maps to ${\bf R}^8$ with coordinates
$ p_{000},p_{001},p_{010},p_{011},p_{100},p_{101},p_{110},p_{111}$.
The map $f : {\bf R}^8 \rightarrow {\bf R}^8$ is given by
\begin{eqnarray*} & f_{\sigma_1 \sigma_2 \sigma_3} \quad
= \quad s_{00} s_{00} t_{0\sigma_1} t_{0\sigma_2} t_{0\sigma_3} +
s_{00} s_{01} t_{0\sigma_1} t_{0\sigma_2} t_{1\sigma_3} +
s_{01} s_{10} t_{0\sigma_1} t_{1\sigma_2} t_{0\sigma_3} +
s_{01} s_{11} t_{0\sigma_1} t_{1\sigma_2} t_{1\sigma_3} \\ &
\phantom{halloween} \qquad
s_{10} s_{00} t_{1\sigma_1} t_{0\sigma_2} t_{0\sigma_3} +
s_{10} s_{01} t_{1\sigma_1} t_{0\sigma_2} t_{1\sigma_3} +
s_{11} s_{10} t_{1\sigma_1} t_{1\sigma_2} t_{0\sigma_3} +
s_{11} s_{11} t_{1\sigma_1} t_{1\sigma_2} t_{1\sigma_3}.
\end{eqnarray*}
The hidden Markov model (i.e. the image of $f$) is the zero set of
the quartic polynomial
\begin{eqnarray*} & p_{011}^2 p_{100}^2-p_{001}^2 p_{110}^2
+ p_{000} p_{011} p_{101}^2-p_{000} p_{101}^2 p_{110} +
p_{000} p_{011} p_{110}^2-p_{001} p_{010}^2 p_{111} +
p_{001}^2 p_{100} p_{111}
   \\ & + p_{010}^2 p_{100} p_{111}
-p_{001} p_{100}^2 p_{111}-p_{000} p_{011}^2 p_{110}
  -p_{001} p_{011} p_{100} p_{101}-p_{010} p_{011} p_{100} p_{101}  \\
  & + p_{001} p_{010} p_{011} p_{110}-p_{010} p_{011} p_{100} p_{110}
  + p_{001} p_{010} p_{101} p_{110} + p_{001} p_{100} p_{101} p_{110}
+ p_{000} p_{010} p_{011} p_{111} \\ & -p_{000} p_{011} p_{100} p_{111}
-p_{000} p_{001} p_{101} p_{111} + p_{000} p_{100} p_{101} p_{111}
+ p_{000} p_{001} p_{110} p_{111}-p_{000} p_{010} p_{110} p_{111}.
  \end{eqnarray*}
  This polynomial was found by a \emph{Gr\"obner basis} computation.
  See the discussion on \emph{implicitization} in \cite[\S 3]{Cox:96}.

In general,
the  polynomial functions on ${\bf R}^{l^n}$ which vanish on the
image of $f$ are the called \emph{invariants of the model}.
They form a prime ideal $I_f$. In our example, $I_f$
is generated by the quartic polynomial above.
Problem 5 is to compute generators of the ideal $I_f$.
When $l^n$ and $d$ are small, this can be done
  using Gr\"obner bases, and in some cases it is
possible to characterize $I_f$ based on the structure of the model
(see, for example, Conjecture \ref{luckydeterminants}), but
in general Problem 5 is hard and
the ideal $I_f$ may remain unknown.

   Here is where tropical geometry comes in.
  The \emph{tropicalization}  of our  map $f$
is the map $g : {\bf R}^d \rightarrow {\bf R}^{l^n}$  defined
by replacing products by sums and sums by minima in the formula for
$f$. In  our example $(n=3,k  =  l  =  2)$,
the tropicalization is the piecewise-linear map
$\,g: {\bf R}^8 \rightarrow {\bf R}^8, \,(U,V) \mapsto \delta \,$ with
\begin{equation} \label{maxformula}
\delta_{\sigma_1 \sigma_2 \sigma_3} \quad
  = \quad \min \,\bigl\{\, u_{h_1 h_2}  + u_{h_2 h_3}
  +  v_{h_1 \sigma_1} +  v_{h_2 \sigma_2} + v_{h_3 \sigma_3} \,\, :
  \,\,(h_1,h_2,h_3) \in \{0,1\}^3 \,\bigr\}.
\end{equation}
This minimum is attained by the  most
  likely hidden data
$(\hat{h}_1,\hat{h}_2,\hat{h}_3)$, given the
observations $(\sigma_1,\sigma_2,\sigma_3)$
and given the parameters
$u_{\cdot \cdot} = - {\rm log}(s_{\cdot \cdot})$ and
$v_{\cdot \cdot} = - {\rm log}(t_{\cdot \cdot})$.
The sequence $(\hat{h}_1,\hat{h}_2,\hat{h}_3)$
is known as the \emph{Viterbi sequence}
in the HMM literature \cite{Rabiner:89}.
 It solves Problem 2 in the Introduction.

The key observation, which we discuss in more detail in
Section 4, is that the set of parameters $(U,V)$ which select
  the Viterbi sequence $\,(\hat{h}_1,\hat{h}_2,\hat{h}_3)\,$
is the normal cone at a vertex of the Newton polytope of
the polynomial $\,f_{\sigma_1 \sigma_2 \sigma_3}$.
This polytope is $4$-dimensional, it has $8$ vertices,
and its normal fan represents the solution to
  Problem 4 in the Introduction when
  $ {\bf \sigma} = \sigma_1 \sigma_2 \sigma_3$
is fixed.

We can also consider an extension of
Problem 4 where $ {\bf \sigma} = \sigma_1 \sigma_2 \sigma_3$
ranges over all possible observations. The solution
is given by the Newton polytope  of the map $f$.
In our example, this is a $5$-dimensional polytope with
$398$ vertices, $1136$ edges, $1150$ two-faces,
$478$ three-faces and $ 68$ facets, namely, the
  Minkowski sum of eight copies of the earlier $4$-dimensional polytope
for $(\sigma_1,\sigma_2,\sigma_3) \in \{0,1\}^3$.
For a concrete numerical example, fix the parameters
$\,U^* = \binom{\,6 \,\,\, 5\,}{\,8 \,\,\, 1 \,}$ and
$\,V^* = \binom{\,0 \,\,\, 8\,}{\,8 \,\,\, 8 \,}$. We find:
$$\begin{matrix}\hbox{if the observed string at} \,\,\,
  Y_1 Y_2 Y_3\, \,\,\hbox{is} &
  \quad \sigma_1 \sigma_2 \sigma_3 \,\, = &
000 & 001 & 010 & 011 & 100 & 101 & 110 & 111
\\\hbox{then the Viterbi sequence at}\,\,
  X_1 X_2 X_3 \,\, \hbox{is}  &
  \quad \hat{h}_1 \hat{h}_2 \hat{h}_3 \,\,=
& 000 & 001 & 000 & 011 & 000 & 111 & 110 & 111
  \end{matrix}$$
The set of all parameters $(U,V) $
leading to the same conclusions as $(U^*,V^*) $
is the cone defined by
\begin{eqnarray*}
& u_{01}-u_{00}+v_{11}-v_{01} \, \leq \, 0 \,
  , \,\,\,u_{10}-u_{11}+v_{00}-v_{10} \, \leq \, 0 \,
  , \,\,\,u_{00}+v_{01}-u_{10}-v_{11} \, \leq \, 0 \,
  , \\& 2 u_{00}+v_{01}-u_{01}-u_{10}-v_{11}  \, \leq \,
  0 \, , \,\,\,2 u_{11}+v_{10}+v_{11}-u_{00}-
u_{01}-v_{00}-v_{01}
\, \leq \, 0 . \,\,\,\end{eqnarray*}
Our solution to the parametric inference problem 
with respect to all observations simultaneously consists
  of $398$ such  cones. The {\em tropical HMM} is
  the union of the images of these cones under the
piecewise-linear map  $g : (U,V) \mapsto \delta$.
This image is a piecewise-linear set of dimension $7$.
The cone  which contains the chosen parameters $(U^*,V^*)$
mapped to a $7$-dimensional cone in the tropical HMM 
(it spans the hyperplane $\,\delta_{010} = \delta_{100} $)
but most of the other $397$ cones are
mapped to lower-dimensional cones by the map $g$.
The question how the number $398$ grows as the length
$n$ increases will be addressed in Corollary 10.

\section{Positivity and Morphisms in Tropical Geometry}

We have seen that a graphical model is the image of
a polynomial map  $f$ from the space of parameters
to the space of joint probability distributions on the observed
random variables. Furthermore, we have seen that the tropicalization
of $f$ arises naturally in solving Problem 4. In this section we 
study the geometry of tropicalization in the more general setting where
$\, f :   {\bf R}^d  \rightarrow {\bf R}^m $ is an arbitrary polynomial 
map.
In statistical applications, it is usually the case that
each coordinate $f_\sigma$ of the map $f$ is a polynomial
with positive coefficients. If this holds then
the polynomial map $f$ is called \emph{positive}.
We say that $f$ is \emph{surjectively positive}
if, in addition, $f$ maps the positive orthant
surjectively onto the positive points in the image,
in symbols,
\begin{equation}
\label{compposi}
f \bigl( {\bf R}_{> 0}^d \bigr) \quad = \quad
  {\rm image}(f) \, \cap \, {\bf R}_{> 0}^m.
\end{equation}
The set of all polynomial functions which vanish
on the image of $f$ is a prime ideal $I_f$ in
the polynomial ring ${\bf R}[p_1,\ldots,p_m]$.
The closure of the image of $f$ is the
variety of the prime ideal $I_f$.

In tropical geometry, we replace the variety of $I_f$
by a piecewise-linear set as follows. The
\emph{tropical variety} ${\cal T}(I_f)$ is the
set of all weight vectors $w \in {\bf R}^m$ such that
the initial ideal ${\rm in}_w(I_f)$ contains no
monomial \cite{Richter-Gebert:03, Sturmfels:96}.
Following \cite{Speyer:03}, we define the
\emph{positive tropical variety} ${\cal T}^+(I_f)$ as the
set of all weight vectors $w \in {\bf R}^m$ such that
the initial ideal ${\rm in}_w(I_f)$ contains no
polynomial with only positive
coefficients. The tropical variety
${\cal T}(I_f)$ is a \emph{polyhedral fan}
in ${\bf R}^m$, and  ${\cal T}^+(I_f)$
is a \emph{polyhedral subcomplex} of ${\cal T}(I_f)$.
This means that ${\cal T}(I_f)$ is a finite union of
closed convex polyhedral cones that fit together nicely,
and ${\cal T}^+(I_f)$ is the union of a subset of these cones.
The \emph{tropicalization}  of the polynomial map $f$
is the piecewise-linear map $g : {\bf R}^d \rightarrow {\bf R}^{m}$  defined
by replacing products by sums and sums by minima in the
evaluation  of $f$. We say that $g$ is a \emph{tropical morphism}.
Examples of tropical morphisms appear in the displayed formulas
(\ref{maxformula}),
(\ref{formula1}),
(\ref{formula2}),
(\ref{formula3}),
(\ref{tropicaltreemap}) and
(\ref{treemapexample3}).

The following theorem describes the geometry of this situation.  We 
define
the \emph{Newton polytope} of a polynomial map
$f : {\bf R}^d \rightarrow  {\bf R}^{m}$  as the Minkowski sum
   in ${\bf R}^d$ of the Newton polytopes of  its
coordinates $f_1,\ldots,f_{m}$.
    For basics on Newton polytopes and their normal fans
see \cite[\S 1]{Sturmfels:96}.

   \begin{thm} \label{hmmmain}  The tropical morphism $g$ is
  linear on each cone in the normal  fan of the
  Newton polytope of  $f$.
Its image is a fan contained in ${\cal T}(I_f)$.
If $f$ is positive then  $\,{\rm image}(g)\,$ is a subset of
${\cal T}^+(I_f)$, but it is generally not a polyhedral subcomplex.
If $f$ is surjectively positive then
  $\,{\rm image}(g)\,= \,{\cal T}^+(I_f)$.
   \end{thm}

\begin{proof}
Let $P_i$ denote the Newton polytope of the polynomial $f_i = f_i(s_1,\ldots,s_d)$.
By definition, $P_i$ is the convex hull in ${\bf R}^d$ of all
non-negative lattice points $a = (a_1,\ldots,a_d) \in {\bf N}^d$ such that the
monomial $s_1^{a_1} \cdots s_d^{a_d}$ appears with non-zero coefficient
in $f_i$. The piecewise-linear concave function $g_i$ is
the \emph{support function} of the polytope $P_i$. This
means that $g_i(w)$ is the minimum value attained on $P_i$
by the linear functional $\,a \mapsto w \cdot a$.
In particular, the function $\,g_i : {\bf R}^d
\rightarrow {\bf R}\,$ is linear on each cone in
the normal fan of $P_i$.

The Newton polytope of the map $f$ is the Minkowski sum
$\,P_1 + \cdots + P_m \,=\,\{ a_1+\cdots+a_m\,:\, a_i \in P_i \}$.
 The normal fan of
$P_1 + \cdots + P_m$ is the common refinement
of the normal fans of $P_1,\ldots,P_m$. This
shows that the function  $\,f = (f_1,\ldots,f_m)
\,:\, {\bf R}^d \rightarrow {\bf R}^d\,$ is linear
on  each cone of the normal fan of
the Newton polytope of $f$. Since $g$ is
continuous,  the image of $g$
is a closed polyhedral fan in ${\bf R}^m$.

Consider any vector $w \in {\bf R}^d$. We
must show that $g(w)$ lies in
${\cal T}(I_f)$, and if $f$ is positive then
$g(w)$ lies in  ${\cal T}^+(I_f)$.
Let $\phi$ be any polynomial in the ideal $I_f$.
If we substitute
$p_1 = f_1, \ldots, p_m = f_m$ into $\phi = \phi(p_1,\ldots,p_m)$
then we get zero.
Consequently, if we substitute
the initial forms $\,p_1 = {\rm in}_{w}(f_1), \ldots,
p_m = {\rm in}_{w}(f_m)\,$ into the initial form
$\,{\rm in}_{g(w)}(\phi)\,$ then the result is zero. See
equation (11.2) on page 100 in
\cite{Sturmfels:96}. This implies that
$\,{\rm in}_{g(w)}(\phi)\,$ is not a monomial. Moreover,
if $f$ is positive then $\phi$ must have two terms
whose coefficients have opposite signs.
This implies the desired
conclusion.

The following example shows that
  ${\rm image}(g)$  need not be a subcomplex
of ${\cal T}^+(I_f)$.
If $f$ is assumed to be surjectively positive,
then it follows from \cite[Proposition 2.5]{Speyer:03}
that $\,{\rm image}(g) = {\cal T}^+(I_f)$.
\end{proof}

\begin{ex} \rm
Let $d = 3$, $m = 4$ and  consider the linear map
$$ f \, : \, {\bf R}^3 \rightarrow {\bf R}^4\, ,\,\,
(s_1,s_2,s_3) \, \mapsto \,
\bigl(\,
s_1+s_2+s_3, \,
s_1+2 s_2+s_3,\,
s_2 + s_3,\,
s_3 \, \bigr). $$
Then $\, I_f\,$ is the
principal ideal generated by the linear form
$\, p_1 - p_2 + p_3 - p_4 $, and
${\cal T}(I_f)$ is essentially the normal fan of a tetrahedron.
We identify ${\cal T}(I_f)$ with the
complete graph $K_4$.  The six edges
of $K_4$ are labeled with six
monomial-free initial ideals of $I_f$, namely,
$$
\langle p_1 +  p_3 \rangle ,\,
\langle - p_2 -  p_4 \rangle ,\,
\langle p_1 - p_2 \rangle ,\,
\langle  p_1  - p_4  \rangle ,\,
\langle - p_2 + p_3 \rangle,\,
\langle  p_3 - p_4  \rangle .$$
The first two of these six initial ideals
contain a polynomial with positive coefficients.
Hence  the positive tropical variety
${\cal T}^+(I_f)$ is the four-cycle
in $K_4$ formed by the remaining
four edges.

The tropicalization of the linear map $f$ is
the tropical morphism
\begin{equation}
\label{formula1}
 g \,: \, {\bf R}^3 \rightarrow {\bf R}^4,
\, (u_1,u_2,u_3) \,\, \mapsto \,\,
\bigl(\,
{\rm min} (u_1,u_2,u_3),\,
{\rm min} (u_1,u_2,u_3),\,
{\rm min} (u_2,u_3),\,u_3 \, \bigr). 
\end{equation}
The image of $g$ is the set of all
vectors $(a,a,b,c)$ with $\, a \leq b \leq c$.
Each vector $(a,a,b,c)$ with $\, a < b < c\,$
has the initial ideal $\langle p_1 - p_2 \rangle$,
so it lies on a particular edge of $K_4$. But
the same edge also accounts for all
vectors $(a,a,b,c)$ with $\, a < c < b$, none
of which is in the image of $g$. Thus $\,{\rm image}(g)$
is a closed segment which covers only half of the
edge of $K_4$ indexed by
$\, \langle p_1 - p_2 \rangle $.

Here it is easy to replace $f$ by a parameterization $f'$
which is surjectively  positive, for instance,
$$ f' \, : \, {\bf R}^4 \rightarrow {\bf R}^4\, ,\,\,
(s_1,s_2,s_3,s_4) \, \mapsto \,
\bigl(\,
s_1+s_3,
s_1+s_4,
s_2 + s_4,\,
s_2+s_3 \, \bigr). $$
\begin{equation}
\label{formula2}
 g' \,: \, {\bf R}^4 \rightarrow {\bf R}^4,
\, (u_1,u_2,u_3,u_4) \,\, \mapsto \,\,
\bigl(\,
{\rm min} (u_1,u_3),\,
{\rm min} (u_1,u_4),\,
{\rm min} (u_2,u_4),
{\rm min} (u_2,u_3) \, \bigr). 
\end{equation}
We have $I_f = I_{f'}$ but now the tropical morphism
$g'$ maps onto the entire four-cycle ${\cal T}^+(I_f)$. \qed
\end{ex}

In the rest of this section we examine
Theorem \ref{hmmmain} for a small but important
graphical model, namely, the \emph{naive Bayes model with
two features}  \cite[\S 7]{Garcia:03}.
There are two observed random variables $Y_1$ and $Y_2$
dependent on one hidden binary random variable $X$.
The two observed variables take $k$ and $l$ possible
values respectively. The parameterization $f$ of this model
is the map $\, f : {\bf R}^{2(k+l)} \mapsto {\bf R}^{kl} \,$
given by $$\,p_{ij}\,\, \, = \,
  \,\,
s_{i0} t_{0j} +  s_{i1} t_{1j} . $$
Thus the model consists of all
$k \times l$-matrices $P = (p_{ij})$
of the form $\,P = S \cdot T\,$
where $S $ is a $k \times 2$-matrix
and $T $ is a $2 \times l$-matrix, i.e.,
 the model consists of precisely
the $k \times l$-matrices of rank $\leq 2$.

\begin{prop} \label{threebythree}
The parameterization $f$ of the naive Bayes model
with two features  is surjectively positive.
The ideal $I_f$ is generated by the
$3 \times 3$-subdeterminants of the
$k \times l$-matrix $P=(p_{ij})$.
\end{prop}

\begin{proof}
The map $f$ being positive means that if
$P$ is any positive matrix of rank $2$ then
$S$ and $T$ can be chosen to be positive.
This is a known result in linear algebra
(see e.g.~\cite{Cohen:93}). The same statement
is false for  rank $\geq 3$, i.e., the
parameterization of the naive Bayes model
with three or more features is not
surjectively positive.
A well-known result in commutative algebra
states that the $(r+1) \times (r+1)$-minors
of a $k \times l$-matrix generate a prime ideal.
The variety of this ideal is the set of $k \times l$-matrices of rank $\leq r$.
This our ideal $I_f$ for $r = 2$.
\end{proof}

The objects of Theorem \ref{hmmmain} have been
studied in \cite{Develin:03} and \cite{Develin:04}.
The tropical variety ${\cal T}(I_f)$ is the
set of $k \times l$-matrices of
\emph{tropical rank} $\leq 2$, and
the tropical variety
$\,{\cal T}^+(I_f) = {\rm image}(g)\,$ is the
set of $k \times l$-matrices of
\emph{Barvinok rank} $\leq 2$.
Develin \cite{Develin:04} determines
the combinatorics and topology of these
spaces when $\,{\rm min}(k,l) = 3$.
He shows that  $\,{\cal T}(I_f)\,$
is shellable  but $\,{\cal T}^+(I_f) \,$ can have torsion in its integral homology groups.

The Newton polytope of the map $f$ is an
interesting combinatorial object, namely, it is
the $(kl-k-l+2)$-dimensional zonotope associated
with the complete bipartite graph $K_{k,l}$.
The Newton polytope of each coordinate 
$f_{ij}$ is a line segment, and the zonotope
is their Minkowski sum.
The normal fan is the hyperplane arrangement 
$\,\{u_{i0} - u_{i1} =  v_{1j} - v_{0j}\}$.
Its maximal cones correspond to the acyclic orientations of the complete bipartite graph $K_{k,l}$.
West \cite{West:95} showed that the number of facets of
such a cone can be any integer between $k+l-1$ and $kl$.
The total number of cones equals
$ \,\sum_{i=1}^{k}S(k,i)(-1)^{l+i}i!(i+1)^{l}$,
where $S(k,i)$ is the Stirling number of the second kind.
Here, the tropical morphism $g$ is given by 
\begin{equation}
\label{formula3}
\,g_{ij} \,\,\, = \, \,\,
{\rm min} \bigl(\,u_{i0}+ v_{0j} ,\,  u_{i1} + v_{1j}\,\bigr) .
\end{equation}
The map $\,g :  {\bf R}^{2(k+l)} \mapsto {\bf R}^{kl} \,$
is piecewise-linear with respect to the hyperplane arrangement.
Recent work of Federico Ardilla (in preparation) gives 
 a complete classification of all fibers of $g$.

\begin{figure}[ht]
\vskip -0.3cm
\begin{center}
  \includegraphics[scale=0.5]{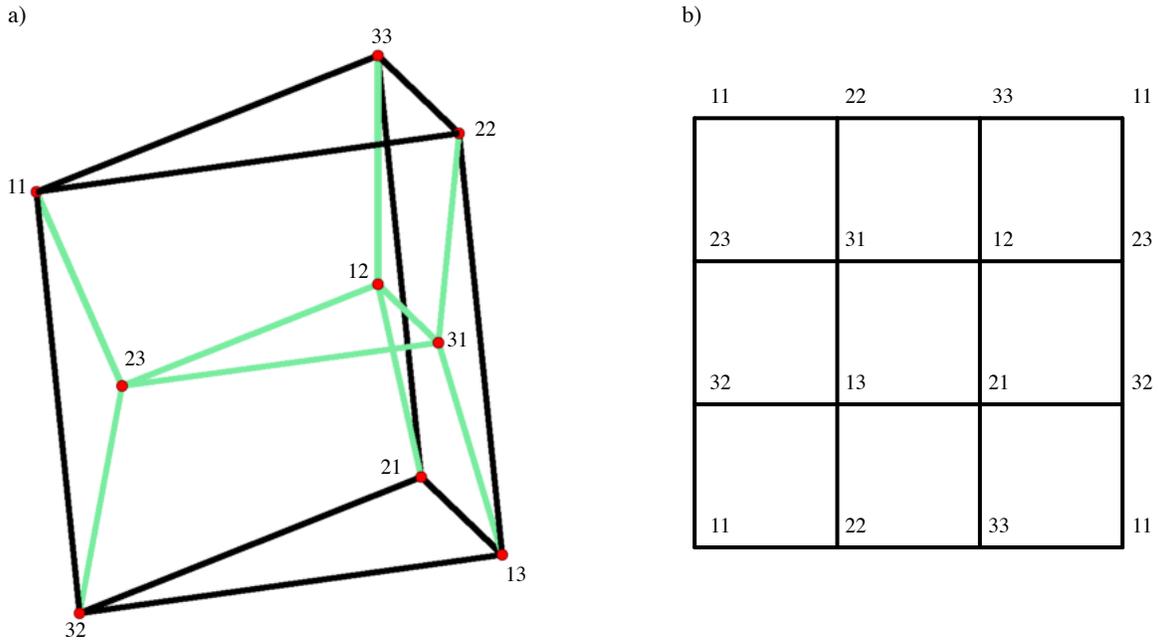}
  \end{center}
\vskip -1.2cm
\caption{The tropical variety and positive tropical
variety of the $3 \times 3$-determinant.}
     \label{fig:naivebayes}\end{figure}

\begin{ex} \rm
Let $k=l=3$, so the two observed random
variables are ternary. The prime ideal is
$$ I_f \quad = \quad
\langle
p_{11}  p_{22}  p_{33}  \, - \,
p_{11}  p_{23}  p_{32}  \, - \,
p_{12}  p_{21}  p_{33}  \, + \,
p_{12}  p_{23}  p_{31}  \, + \,
p_{13}  p_{21} p_{32}   \, - \,
p_{13}  p_{22} p_{31} \rangle.
$$
The tropical variety ${\cal T}(I_f)$ is the fan over a
two-dimensional polyhedral complex consisting of
  six triangles and nine quadrangles. This complex
is the $2$-skeleton of the product of two triangles,
labeled as in Figure 2a.
This complex is shellable.
The positive tropical variety ${\cal T}^+(I_f)$ is
the subcomplex consisting of the nine quadrangles shown in
Figure 2b. Note that ${\cal T}^+(I_f)$ is
a torus.

The Newton polytope of $f$ is a five-dimensional
zonotope with $230$ vertices, one for each
acyclic orientation of the complete bipartite graph $K_{3,3}$. The map $g$
is linear on each of the $230$ cones in the
corresponding hyperplane arrangement,  but it is
rank-deficient on $68$ of the cones.
The remaining $\,162 = 18 \times 9 \,$ cones are mapped onto the
$9$ quadrangles of the torus ${\cal T}^+(I_f)$.
Thus the general fiber of $g$ involves $18$ cones.
Of these, eight cones  have  $5$ facets,
eight cones have $6$ facets,
and two cones have $9$ facets. \qed
\end{ex}

\section{Newton Polytopes of Graphical Models and their Complexity}

Consider a graphical model with $E$ edges and  $n$ observed random 
variables
$Y_1,\ldots,Y_n$ each taking $l$ values. Such a model is
given by a positive polynomial map
$\, f \, : \, {\bf R}^d \rightarrow {\bf R}^{l^n}$.
Each coordinate $f_{\bf \sigma}$ of $f$ is a polynomial of
degree $e$ in the model parameters $s_1,\ldots,s_d$.
In this section we discuss the statistical meaning
and the computational complexity of the mathematical objects
introduced in the previous section.

We write $u_i = - {\rm log}(s_i)$
for the negative logarithms of the model parameters.
Consider any of the $l^n$ possible observations
${\bf \sigma}$.
The quantity $\,f_\sigma(s_1,\ldots,s_d)$ is the
probability of making this particular observation, i.e.~it is ${\rm Prob}({\bf Y} = {\bf \sigma})$.
The quantity $\,g_\sigma(u_1,\ldots,u_d)\,$
is the negative logarithm of the conditional probability
$ {\rm Prob}({\bf X}=\widehat {\bf h} \, |\,{\bf Y} = {\bf \sigma})$ where $\widehat {\bf h}$
maximizes $\,{\rm Prob}({\bf X}= {\bf h}\, |\, {\bf Y} = {\bf \sigma})$ 
for the parameters $(s_1,\ldots,s_d)$.
Clearly, the function $\, g_\sigma :{\bf R}^d \rightarrow {\bf R}  \,$
is piecewise-linear and concave  on the logarithmic parameter space.

The domains of linearity of the function $g_{\bf \sigma}$
are the cones in the normal fan of the
Newton polytope of $f_{\bf \sigma}$. Each maximal cone
$\cal{C}$ is indexed by the hidden data $\widehat {\bf h}$
that maximizes ${\rm Prob}({\bf X}={\bf h}|{\bf Y}={\bf \sigma})$
for any of the parameters $(u_1,\ldots,u_d) \in {\cal C}$.
The hidden data   $\widehat {\bf h}$
which arise in this manner,
for some choice of logarithmic parameters $ u$,
are called the possible \emph{explanations}
of the observation ${\bf \sigma}$. 
For instance, for the hidden
Markov model of Section 2, the
explanations are the Viterbi sequences.

Let us now vary the observations.
Each logarithmic parameter vector ${\bf u}$ defines an
\emph{inference function}  ${\bf \sigma} \mapsto \widehat {\bf h}$
from the set of observations to the set of
explanations. For the HMM, each inference function
$\,\{1,\ldots,l\}^n \rightarrow \{1,\ldots,k\}^n \,$
takes an observed sequence $\sigma$
to the corresponding Viterbi sequence $\widehat {\bf h}$.
There are $\,(k^n)^{l^n} \, = \, k^{nl^n}\,$ such functions, but most
of these are {\bf not} inference functions. For instance,
consider the binary HMM of length three.  There are
$\, 8^8 = 16,777,216 \,$ Boolean functions
$\,\{0,1\}^3 \rightarrow \{0,1\}^3  $, but,
as we have seen at the end of Section 2, only $398$ of these are
inference functions for the HMM.

\begin{prop}
The inference functions ${\bf \sigma} \mapsto \widehat {\bf h}$
of a graphical model $f$ are in bijection with the
vertices of the Newton polytope of the map $f$.
The  explanations $ \widehat {\bf h} $ for a fixed observation  ${\bf \sigma}$
in a graphical model are in bijection with the
vertices of the Newton polytope of the polynomial $f_{\bf \sigma}$.
\end{prop}

In applications of graphical models,
the number $d$ of parameters and
the number $l$ of values of the observed
random variables is small and fixed, but
the number $n$ of observed random variables
is large. Recall that the
model is the image of the map
$\, f \, : \, {\bf R}^d \rightarrow {\bf R}^{l^n}$.
Hence the dimension of the model remains fixed
but the dimension of its ambient space
grows exponentially in $n$. It is therefore algorithmically infeasible
to compute the full tropical variety ${\cal T}(I_f)$.
What we can do efficiently, however, is to compute
the Newton polytopes of the $f_{\bf \sigma}$,
or even the Newton polytope of  $f$. This allows us to glean
information about the tropical variety
from the domains of linearity of its
``coordinate functions'' $g_{\bf \sigma}$.

Our next goal is to derive an upper bound
on the number of vertices of the Newton polytopes.

\begin{thm} \label{smallpolytopebound}
Consider graphical models $f$ whose number of parameters $d$ is fixed
and whose number $n$ of observed random variables and number of edges $E$ varies.
(Typically, $E$ is a linear function of $n$).
Then the number of vertices of the
Newton polytope $NP(f_\sigma)$ of  $f_\sigma$ is bounded above by
$$ \# \,{\rm vertices} (NP(f_\sigma)) \,\,\, \leq \,\,\,
  {\rm constant} \cdot E^{d(d-1)/(d+1)} \,\,\, \leq \,\,\, {\rm 
constant} \cdot E^{d-1} . $$
\end{thm}

For many important families of graphical models,
the number $E$ of edges is bounded by a linear function
in terms of the number $n$ of observed nodes, and in those cases we can replace $E$ by $n$.
Hence, for any given observation $\sigma$, the number
of explanations grows polynomially in $n$.
For instance, in the hidden Markov model
of Section 2 we have $E = 2n-1$, and a similar relationship
holds in the tree model of Section 5.

\begin{cor}
\label{HMMcomplexity}
For any fixed observation in the homogeneous HMM,
the number of explanations is at most $ \,C_{k,l} \cdot n^{k(k+l)}$. If
all random variables are binary then the upper bound
$\,\,C \cdot n^{10/3} \,\,$ holds.
\end{cor}

The proof of  Theorem \ref{smallpolytopebound}
and Corollary \ref{HMMcomplexity}
are derived from the following
  classical result on lattice polytopes due to
Andrews \cite{Andrews:63}.  The necessary observation is that the
Newton polytope of $f_\sigma$ is contained in the cube
$[0,E]^d$, and the volume of this cube equals $E^d$.

\begin{prop} {\rm (Andrews \cite{Andrews:63}) }
For every fixed integer $d$ there exists a constant $C_d$ such
that  the number of vertices of any lattice polytope $P$ in ${\bf R}^d$
is bounded above by $\, C_d \cdot {\rm volume}(P)^{(d-1)/(d+1)}$.
\end{prop}

The Newton polytope of the map $f$ was defined as the
Minkowski sum of the $l^n$ smaller Newton polytopes
in   Theorem \ref{smallpolytopebound}. From this we infer the following naive bound
on its number of vertices.

\begin{cor}
The number of inference functions of a graphical model
is at most $\, l^{n C_d E^{d-1}} $, hence this number  scales
at most singly exponentially in the complexity $(n,E) $ of the 
graphical model.
\end{cor}

Consider the homogeneous HMM on binary random variables.
Each inference function is a Boolean function
$\,\{0,1\}^n \rightarrow \{0,1\}^n$, but not conversely.
The number of all Boolean functions is $\, 2^{n 2^n}$, which grows
doubly exponentially in $n$. However, the number of
inference functions is at most $\, 2^{{\rm polynomial}(n)}$.

In practical applications of graphical models,
it may be infeasible to compute all (singly-exponentially many) 
inference functions.
Nonetheless, we believe that important insight can be
gained by computing and classifying the Newton polytopes
of graphical models $f$ on few random variables.
Such a study would be the polyhedral analogue to
the algebraic classification of \cite{Garcia:03}.

On the other hand, for a fixed observation $\sigma$, 
the size of the Newton polytope of $f_{\bf \sigma}$ grows polynomially 
with the size of the graphical model, and therefore there is hope 
that the polytopes can be computed efficiently.
Despite the fact that the Newton polytope of $f_{\sigma}$ has polynomially many vertices in the size of the graphical model, the number of terms in $f_{\sigma}$ grows exponentially. This is a potential problem because the 
computation of the Newton polytope requires
inspecting these terms. The following result states that, in fact, 
the convex hull computations scales with the running time of 
the sum-product algorithm, which for many models of interest scales 
polynomially with the size of the graphical model.

\begin{prop}[{\bf Polytope propagation}] \label{polyprop}
The Newton polytopes of the polynomials
$f_\sigma$ can be computed recursively using the decomposition
of $f_\sigma$ according to the sum-product algorithm.
\end{prop}

Taken together, Theorem 7 and Proposition 11 say that {\bf polytope propagation is an efficient algorithm for parametric inference with graphical models}.
This statement is thesis (c) in our companion paper \cite{Pachter:04}. In that paper,
the sum-product algorithm and the polytope propagation algorithm
are explained and analyzed in more detail. We also demonstrate
the practicality of our mathematical theory by explicitly computing
(and statistically interpreting)
various high-dimensional Newton polytopes for graphical models that
arise in biological sequence analysis.

\section{The General Markov Model on a Binary Tree}

We conclude by illustrating the concepts we have developed in the 
context of tree Markov models. These are directed graphical models 
where
the graph is a directed tree $\tau$ with observed random variables
$Y_1,\ldots,Y_n$ at the leaves. The naive Bayes model in Section 3 is 
the special case where $n=2$. Each edge $e$ has
a different transition matrix $S^e = [ s^e_{\mu \nu}]$.
We consider the general model in
Allman and Rhodes \cite{Allman:03},
which means that the $S^e$ are arbitrary
distinct $l\times l$-matrices.
In most applications,
the transition matrices are from a special
model family (e.g. in phylogenetics these may be Jukes-Cantor model 
or the Hasegawa-Kishino-Yano model).  As before, we relax the hypothesis
  that transition probabilities
are non-negative and sum to $1$. Hence the
$s^e_{\mu \nu}$ are distinct unknowns. For simplicity
we shall further assume that the tree $\tau$ is binary.

\begin{prop}
\label{treeobservation}
  The general Markov model for the binary tree $\tau$
is the image of a  map $\, f : {\bf R}^{(2n-2)l^2} \rightarrow {\bf 
R}^{l^n}$,
where each coordinate of $f$ is a multilinear polynomial
in the unknowns $\bigl\{(s^e_{\mu \nu})$, $e$ edge of $\tau 
\bigr\}$.
\end{prop}

\begin{figure}[ht]

\begin{center}
  \includegraphics[scale=0.7]{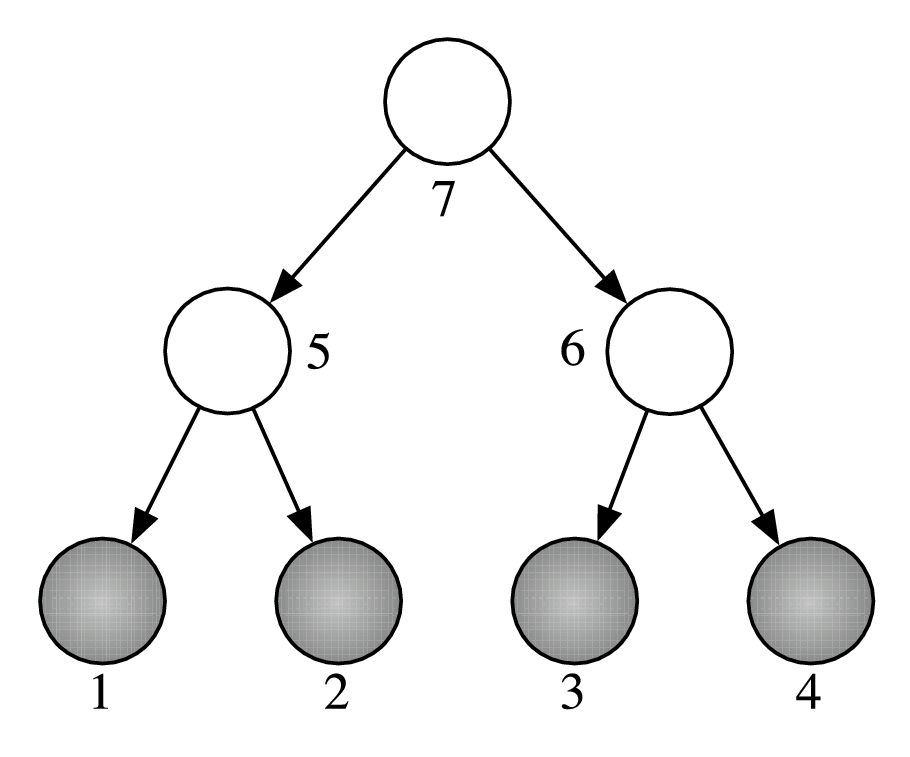}
  \end{center}
\vskip  -0.4cm
\caption{A directed binary tree with $n = 4$ leaves.}
     \label{fig:tree}\end{figure}

If we denote an edge between nodes $i$ and $j$ by $(ij)$
and $\tau'$ is the tree $\tau$ without the leaves, then
the coordinate of the multilinear map $f$ indexed by an
observed sequence 
$(\sigma_1 ,\ldots,\sigma_n)$ can be
written as follows:
\begin{equation}
\label{treemap}
  p_{\sigma_1 \cdots \sigma_n} \quad = \quad \sum_h
\prod_{i \in \tau' \atop
{\rm with}\, {\rm children}\,j,k}
\bigl( s^{(ij)}_{h_i h_j}
    \cdot s^{(ik)}_{h_i h_k} \bigr).
\end{equation}
Here $h$ ranges over all
colorations $\,h = (h_i)_{i \in \tau}\,$ of the nodes
such that $h_j = \sigma_j$ for all leaves $j$.
Our running example in this section is the
binary tree in Figure \ref{fig:tree}
with binary random variables $(l=2)$.

In this example, the coordinates of the
multilinear map $\,f : {\bf R}^{24} \rightarrow {\bf R}^{16} \,$
are given by the formula
\begin{equation}
\label{treemapexample}
p_{\sigma_1 \sigma_2 \sigma_3 \sigma_4} \quad
= \quad
\sum_{\{h_5,h_6,h_7\} \in \{0,1\}^3} \,
(s_{h_7h_5}^{(75)}
\cdot s_{h_7h_6}^{(76)})\cdot
  (s_{h_5\sigma_1}^{(51)}\cdot s_{h_5\sigma_2}^{(52)})\cdot
  (s_{h_6\sigma_3}^{(63)}\cdot s_{h_6\sigma_4}^{(64)}) .
\end{equation}
The  prime ideal $I_f$ of polynomial invariants is generated by the
$3 \times 3$-subdeterminants of the matrix
\begin{equation}
\label{fourbyfour}
\begin{pmatrix}
p_{0000} & p_{0010} & p_{0001} & p_{0011} \\
p_{0100} & p_{0110} & p_{0101} & p_{0111} \\
p_{1000} & p_{1010} & p_{1001} & p_{1011} \\
p_{1100} & p_{1110} & p_{1101} & p_{1111}
\end{pmatrix}
\end{equation}
Thus this particular model is the $k=l=4$ instance of the
determinantal variety in Proposition \ref{threebythree}.

We generalize the determinantal presentation
in this example by proposing the following
explicit solution to Problem 5 for
arbitrary binary trees $\tau$.  Every edge of
$\tau$ induces a \emph{split} of the set of leaves
$\{1,2,\ldots,n\}$, corresponding to
the two connected components of the
tree obtained by removing that edge.
The unrooted tree underlying $\tau$
is uniquely determined by the set of
these splits.

\begin{conj}
\label{luckydeterminants}
The ideal $I_f$ of phylogenetic invariants
of the general Markov model for any binary tree $\tau$ on
binary random variables is generated
by the $3 \times 3$-determinants of all two-dimensional
matrices obtained by \emph{flattening} the
$2 \times \cdots \times 2$-table
$(p_{\sigma_1 \cdots \sigma_n})$ according to
the splits induced by the edges of $\tau$.
\end{conj}

We need to explain the meaning of the word ``flattening''.
If $(A , B)$ is any split of the set $ \{1,\ldots,n\}$
then this refers to the $\,2^{\#(A)} \times 2^{\#(B)}$-matrix
whose rows and columns are indexed by functions $\, A \rightarrow 
\{0,1\}\,$
and $\, B \rightarrow \{0,1\}\,$ respectively,
and whose entries are the $2^n$ probabilities $p_{\sigma_1 \cdots \sigma_n}$.

In December 2003, Allman and Rhodes announced a proof
of the set-theoretic version of our Conjecture
\ref{luckydeterminants}. What this means algebraically
is that $I_f$ equals the  radical  of the ideal generated by
the aforementioned $3 \times 3$-determinants. In light
of this progress, we wish to offer also the following
tropical version of  Conjecture  \ref{luckydeterminants}.
It would be very nice to show that Proposition \ref{threebythree}
 extends to this situation.
However, none of the remaining discussion in this
section depends on these conjectures.

\begin{conj}
\label{luckytropical}
The map $f$ is surjectively positive for $l=2$.
The tropical variety (resp.~the positive tropical variety)
of the prime ideal $I_f$ coincides with the set of
all $2 \times 2 \times \cdots \times 2$-tables
$(u_{\sigma_1 \cdots \sigma_n})$
whose flattenings along the splits
of the tree $\tau$ have tropical rank
(resp.~Barvinok rank) at most $2$. \end{conj}

The sum-product algorithm is used in practice to evaluate
the polynomial  (\ref{treemap}).
Its running time is linear in $n$, despite the
fact that the number $l^{n-1}$ of terms in (\ref{treemap})
  grows exponentially.  This reduction in complexity is achieved
by recursively grouping subsums.
For instance,  (\ref{treemapexample}) becomes

\begin{equation}
\label{treemapexample2}
p_{\sigma_1 \sigma_2 \sigma_3 \sigma_4} \quad = \quad
\sum_{\nu=0}^1 \, \bigl(
s_{\nu0}^{(75)}s_{0\sigma_1}^{(51)}s_{0\sigma_2}^{(52)}
  +s_{\nu1}^{(75)}s_{1\sigma_1}^{(51)}s_{1\sigma_2}^{(52)} \bigr) \cdot
  \bigl( s_{\nu0}^{(76)}s_{0\sigma_3}^{(63)}s_{0\sigma_4}^{(64)}
  +s_{\nu1}^{(76)}s_{1\sigma_3}^{(63)}s_{1\sigma_4}^{(64)} \bigr) .
\end{equation}
The rule to remember is this:
Polynomials are evaluated
recursively as sums of products of smaller
polynomials. This is the solution to Problem 1.
For details on the 
tree case see
   \cite{Durbin:98}.

Problem 2 is known in phylogeny as the
\emph{joint ancestral reconstruction} problem, which asks
for the maximum a posteriori
ancestral assignments $\hat h_i$ given the
observations $(\sigma_1,\ldots,\sigma_n)$
at the leaves. An efficient method for
solving this problem appears in \cite{Pupko:00}.
This method is nothing but the  sum-product algorithm 
with ordinary arithmetic $(+, \times)$ replaced
by tropical arithmetic $({\rm min}, +)$.
The $\sigma$-coordinate of the tropicalization
$\, g : {\bf R}^{(2n-2)l^2} \rightarrow {\bf R}^{l^n}\,$
 of the map (\ref{treemap}) is
\begin{equation}
\label{tropicaltreemap}
  \delta_{\sigma_1 \cdots \sigma_n} \quad = \qquad \min_h \
\sum_{i \in \tau' \atop
{\rm with}\, {\rm children}\,j,k}
\bigl( v^{(ij)}_{h_i h_j} +  v^{(ik)}_{h_i h_k} \bigr),
\end{equation}
This expression can be evaluated efficiently
by the same scheme as before. The rule now is this:
Piecewise-linear concave functions are evaluated
recursively as minima of sums of smaller such functions.
A simple example illustrating this rule is the
  tropicalization of
(\ref{treemapexample2}):
\begin{equation}
\label{treemapexample3}
\delta_{\sigma_1 \sigma_2 \sigma_3 \sigma_4} \quad = \quad
\min_{\nu \in\{0,1\}} \,(
u_{\nu \sigma_1 \sigma_2}\, + \, u_{\nu \sigma_3 \sigma_4})
\end{equation}
where  $\,u_{\nu \sigma_1 \sigma_2}\, =
\min
\bigl( v^{(75)}_{\nu 0} + v^{(51)}_{0 \sigma_1} + v^{(52)}_{0 \sigma_2} 
  ,\,
  v^{(75)}_{\nu 1} + v^{(51)}_{1 \sigma_1} 
+ v^{(52)}_{1 \sigma_2} \bigr)\,$
and similarly for $\,u_{\nu \sigma_3 \sigma_4}$.

We saw in Section 4 that the number of vertices of the Newton polytopes
of the coordinate polynomials $f_\sigma$ is  critical for
efficient parametric inference. That number grows
polynomially in $n$ if the number of parameters is fixed
(thanks to Theorem \ref{smallpolytopebound}) but it may grow
exponentially if the number of parameters is not bounded.
For the general Markov model on a tree $\tau$, the  growth will be 
exponential unless we restrict the number of parameters. This
can be done, for instance, by considering the
\emph{homogeneous tree model} where the
transition matrices along all edges are identical:
$$ s^{e}_{\mu \nu} \,\, = \,\, s_{\mu \nu} \qquad
\hbox{is independent of the edge $e$}. $$
Using Theorem \ref{smallpolytopebound}, we obtain
the following result analogous to Corollary  \ref{HMMcomplexity}.

\begin{prop}
The number of vertices of the Newton polytope of any coordinate $f_\sigma$
in the homogeneous tree model is bounded above by
$n^{l^2-1}$ times a constant depending only on $l$.
\end{prop}

For  tree models which are used in applications, such as phylogenetics,
 the number of  parameters is likely to be reduced even further.
In such cases, the  parametric
joint ancestral reconstruction problem can be solved efficiently
using the polytope propagation algorithm techniques in 
Proposition \ref{polyprop}.

\section{Summary: A  Statistics -- Geometry  Dictionary}

The algebraic representation for graphical models with hidden variables leads
naturally to an interpretation of a parameterized model as a point on an
algebraic variety. Marginal probabilities are coordinates of points on the variety. 
Varieties can be tropicalized, and the statistical meaning is that the MAP probabilities (calculated with logarithms of the parameters) can be interpreted as coordinates of points on 
the positive part of the tropical variety. Hence, the tropical
model is fundamental for understanding MAP probabilities. 
Although we have not addressed it in this paper, the logarithms of the marginal probabilities are
coordinates of points on the \emph{amoeba} \cite{Viro:02} of the model. Amoebas are likely
to be important for understanding the geometry of maximum likelihood estimation.

The sum-product algorithm for graphical models is an efficient method for 
evaluating the coordinate polynomials of a graphical model. This
algorithm works in exactly the same way for classical arithmetic $(+, \times)$
and for  tropical arithmetic $(+,{\rm min})$. This means that
the same method is used to evaluate coordinates of points on the 
variety and of points on the tropical variety.

An explanation for an observation ${\bf \sigma}$ is a vertex of the Newton polytope of 
$f_{\bf \sigma}$. Thus, the parametric inference problem is solved by finding the normal fans of 
the Newton polytopes of the coordinate polynomials. For many important applications, the number of vertices of the polytopes is polynomial in the size of the graphical model. 
The polytope propagation algorithm, which is a geometric analog of the sum-product algorithm, finds the Newton polytopes, and is efficient when the sum-product algorithm is fast and the number of vertices on the Newton polytopes is small.

An inference function for a graphical model is a function from
the set of observations to the set of explanations which 
maximizes the a posteriori probabilities with respect to some choice of parameters.
Inference functions correspond to vertices of the Newton polytope 
of the map $f$. This polytope is much larger than the Newton polytope
of a single coordinate $f_\sigma$, so it can only be computed for
small graphical models, but it has the advantage that it encodes 
the entire piecewise-linear geometry of the model.

 In a companion paper \cite{Pachter:04}, we show that polytope propagation is practical and useful in the important application of biological sequence analysis. In particular, existing parametric alignment methods \cite{Fernandez-Baca:00, Gusfield:94, Waterman:92} can be viewed as special cases of parametric inference for pair hidden Markov models. The computation of the Newton polytopes 
is also useful for Bayesian computations, where we have priors on the parameters and it is of interest to integrate over the maximal cones in the normal fan of the Newton polytope
\cite[\S 5]{Pachter:04}.

\section{Acknowledgments}
Lior Pachter was supported in part by a grant from the NIH (R01-HG02362-02).
Bernd Sturmfels was supported by
a Hewlett Packard Visiting Research Professorship 2003/2004
at MSRI Berkeley and in part  by the NSF (DMS-0200729).
We are grateful to Komei Fukuda, Michael Joswig and
Kristian Ranestad for their help in obtaining the
computational results reported in Section 2.

\end{document}